\documentclass[twocolumn,
    showpacs,
    preprintnumbers,
    amsmath,
    amssymb,
    longbibliography,
    tightenlines,
    nobibnotes,
    aps,
    floatfix,
    prl]{revtex4-1}
\usepackage{amsfonts}
\usepackage[english]{babel}
\usepackage[T1]{fontenc}
\usepackage{times}
\usepackage{mathrsfs}
\usepackage{graphicx}
\usepackage{dcolumn}
\usepackage{bm}
\usepackage{braket}
\usepackage[colorlinks,bookmarks=true,citecolor=blue,linkcolor=red,urlcolor=blue]{hyperref}
\usepackage[tight, FIGTOPCAP, hang, raggedright, nooneline]{subfigure}

\newcommand{\mean}[1]{\langle #1 \rangle}

\renewcommand{\vec}[1]{\boldsymbol{\mathbf{#1}}}

\usepackage{xcolor}

\renewcommand{\d}[1]{\textrm{d}#1}

\begin{document}
\title{Charge density wave instabilities of type-II Weyl semimetals in a strong magnetic field}

\author{Maximilian Trescher$^1$, Emil J. Bergholtz$^{2}$, Masafumi Udagawa$^{3}$, Johannes  Knolle$^{4}$}
\affiliation{$^1$ Dahlem Center for Complex Quantum Systems and Institut f\"ur Theoretische Physik, Freie Universit\"at Berlin, Arnimallee 14, 14195 Berlin, Germany
\\ $^2$ Department of Physics, Stockholm University, AlbaNova University Center, 106 91 Stockholm, Sweden
\\ $^3$ Department of Physics, Gakushuin University, Mejiro, Toshima-ku, Tokyo 171-8588, Japan
\\ $^4$ T.C.M. group, Cavendish Laboratory, J. J. Thomson Avenue, Cambridge, CB3 0HE, United Kingdom} 
\date{\today}

\begin{abstract}
Shortly after the discovery of Weyl semimetals properties related to the topology of their bulk band structure have been observed, e.g. signatures of the chiral anomaly and Fermi arc surface states. These essentially single particle phenomena are well understood but whether interesting many-body effects due to interactions arise in Weyl systems remains  much less explored. Here, we investigate the effect of interactions in a microscopic model of a type-II Weyl semimetal in a strong magnetic field. We identify a charge density wave (CDW) instability even for weak interactions stemming from the emergent nesting properties of the type-II Weyl Landau level dispersion. We map out the dependence of this CDW on magnetic field strength. Remarkably, as a function of decreasing temperature a cascade of CDW transitions emerges and we predict characteristic signatures for experiments.
\end{abstract}

\maketitle

{\it Introduction.}
The theory of Weyl fermions in condensed matter systems --- semimetals where conduction and valence bands have non-degenerate touching points --- has a long and intriguing history, bringing ideas originally developed in the context of particle physics into the realm of materials physics.\cite{weyl_elektron_1929,volovik_universe_2009, murakami_phase_2007,wan_topological_2011,burkov_weyl_2011} Recently the experimental discovery of a Weyl semimetal was reported by various groups \cite{xu_discovery-taas_2015,lv_discovery_2015,lu_experimental_2015} soon followed by transport measurements demonstrating the chiral anomaly \cite{huang_observation_2015,zhang_signatures_2016} as well as the connection of Fermi arc surface states to chiral bulk modes in strong magnetic fields\cite{Moll2016}. 

Much recent attention has focused on type-II Weyl semimetals \cite{soluyanov_type-ii_2015} whose linear dispersion is so strongly tilted \cite{bergholtz_topology_2015,xu_structured_2015,goerbig_tilted_2008,trescher_quantum_2015} that it forms electron and hole pockets. Several materials in this class, including WTe$_2$ \cite{ali_large_magnetoresistance_wte2_2014}, WP$_2$ \cite{Autes2016,Kumar2017} and  Mo$_x$W$_{1-x}$Te$_2$ \cite{belopolski_discovery_2016}, have in parallel attracted ample attention due to their remarkable magnetotransport properties. 
While a single particle analysis reveals a novel twist on the chiral anomaly in type-II Weyl semimetals \cite{Udagawa2017,Tchoumakov2017,Yu2017}, there is no commonly accepted explanation of the observed magnetotransport properties nor an understanding of whether they are at all rooted in the topological properties of these materials, highlighting the need for an understanding of many-body effects. 

Exotic interaction effects in Weyl semimetals \cite{Meng2016,bergholtz_topology_2015}, some taking place only in systems with tilted Weyl cones \cite{bergholtz_topology_2015}, have recently been explored. More conventional phenomena such as density wave instabilities have so far only been discussed in type-I Weyl semimetals where, however, they require a significant critical interaction strength at zero magnetic field \cite{Wei2012,Wang2013,Laubach2016,Roy2017}, consider the chemical potential away from the Weyl node\cite{wang_topological_2016} or only appear in a magnetic field as effective one-dimensional instabilities of the chiral mode \cite{Yang2011,Roy2015,zhang_transport_2017}.
Further spin ordering in Weyl semimetals has been studied~\cite{sun_helical_2015}.

Here we show that the electron- and holelike pockets of the overtilted cones in type-II Weyl semimetals generically render these systems much more susceptible to interaction effects. In particular we show that in a magnetic field the Landau level dispersion acquires nestinglike features between a large number of Landau level bands which triggers a charge density wave (CDW) transition for small interactions. As this emergent weak coupling instability in a magnetic field neither requires perfect particle-hole compensation nor nesting of the zero field band structure we argue that CDW phases are a common property of the high field regime of type-II Weyl semimetals. 

Our starting point is a simple microscopic Hamiltonian $H=H_0+H_{\text{int}}$ with a quadratic lattice model $H_0$ featuring Weyl nodes and a local density-density interaction $H_{\text{int}}$. We show that in a field a weak coupling intracone CDW instability with a wavevector related to the electron and hole pocket separation appears. Our qualitative discussion is corroborated by a microscopic calculation  for which we derive a continuum description. Going beyond lowest order in the momenta is necessary to describe closed electron  and hole pockets. We note that this is in general crucial for a correct low energy description in type-II Weyl systems. 

This Rapid Communication is structured as follows:
First we introduce a simple microscopic model of a type-II Weyl semimetal. Calculating the dispersion of Landau levels in a magnetic field both in the lattice and in the corresponding continuum description, we give an intuitive argument for a CDW instability based on emergent nesting features.
Second, we develop the mean-field theory for generic interactions in a type-II Weyl cone in a magnetic field.
Third, we present self-consistent CDW solutions as a function of temperature and magnetic field. Finally, we discuss implications for experiments.

{\it The model.}  We first concentrate on the non-interacting bandstructure governed by
\begin{align}
H_0({\mathbf k}) =& \left(M-\cos k_x-\cos k_y\right)\sigma^x +\sin k_y\sigma^y  \nonumber\\
 & + \sin k_z\sigma^z + \left( t_1 \sin k_z+t_2 \sin 2k_z \right)\sigma^0
\label{Lattice_Hamiltonian}
\end{align}
in which the tilt of the Weyl cones can be easily tuned to feature electron and hole pockets, e.g. at $M=1$, $t_1+2t_2>1$ this model has type-II Weyl nodes at $\mathbf k_W=(\pm \pi/2,0,0)$. Since we are interested in intracone instabilities we can expand around one of the  cones yielding a low energy  continuum description
\begin{align}
H_0^{{\rm eff}}({\mathbf k})= &\pm k_x\sigma^x + k_y\sigma^y+(k_z-\frac 1 6 k_z^3)\sigma^z  \nonumber\\
 &+ \left((t_1+2t_2) k_z-\frac 1 6(t_1+8t_2)k_z^3\right)\sigma^0.
\label{Eff_Hamiltonian}
\end{align}
The momenta $k_i$ ($i=x,y,z$) are  in the range $-\pi < k_i \leq \pi$ and measured in $1/a_0$, where $a_0$ is the lattice spacing. We have omitted an overall prefactor $\hbar v$ which sets the energy scale in terms of Fermi velocity $v$. Importantly, we have expanded up to $O(k_x^2,k_y^2,k_z^5)$ and included the next-to-leading-order term in $k_z$. We are not aware that this has been done before but it is crucial for a correct low-energy description with closed electron and hole pockets. 

\begin{figure}[t]
    \centering
    \includegraphics[width=0.43\textwidth]{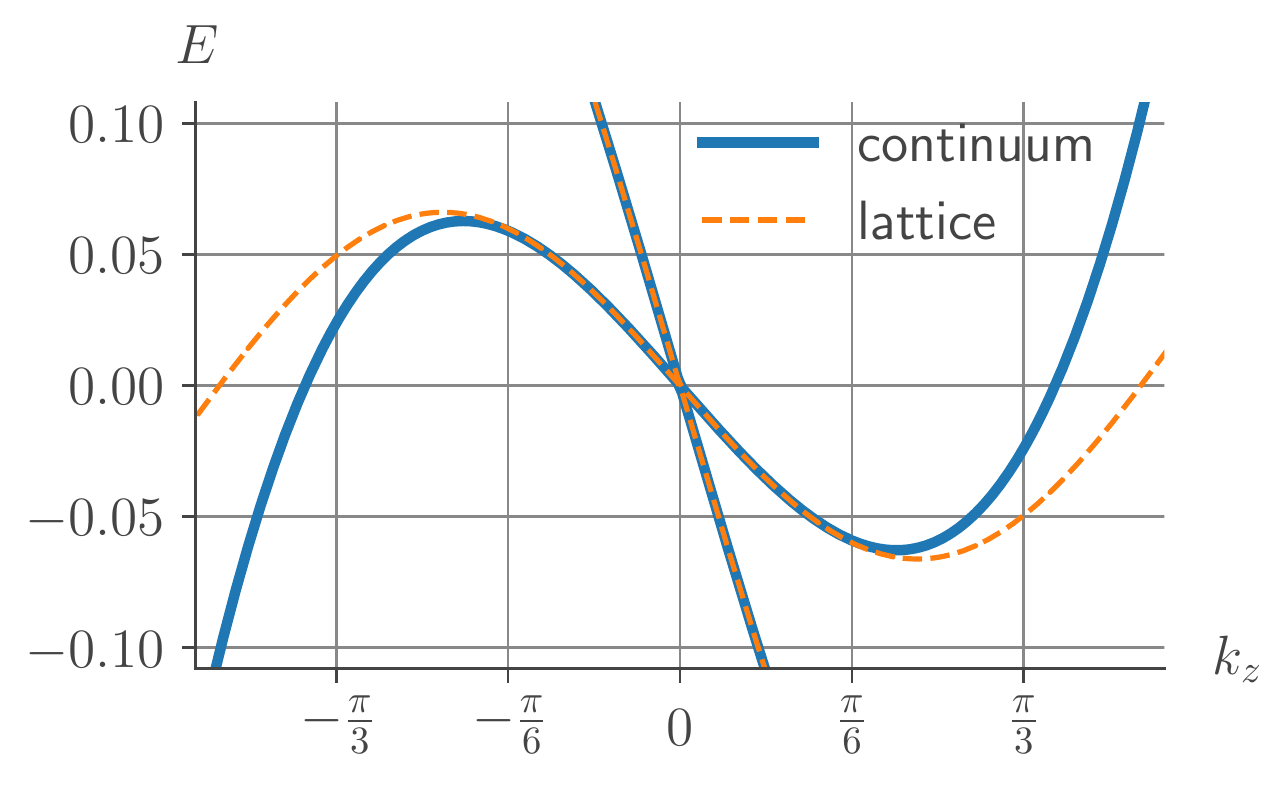}
    \includegraphics[width=0.43\textwidth]{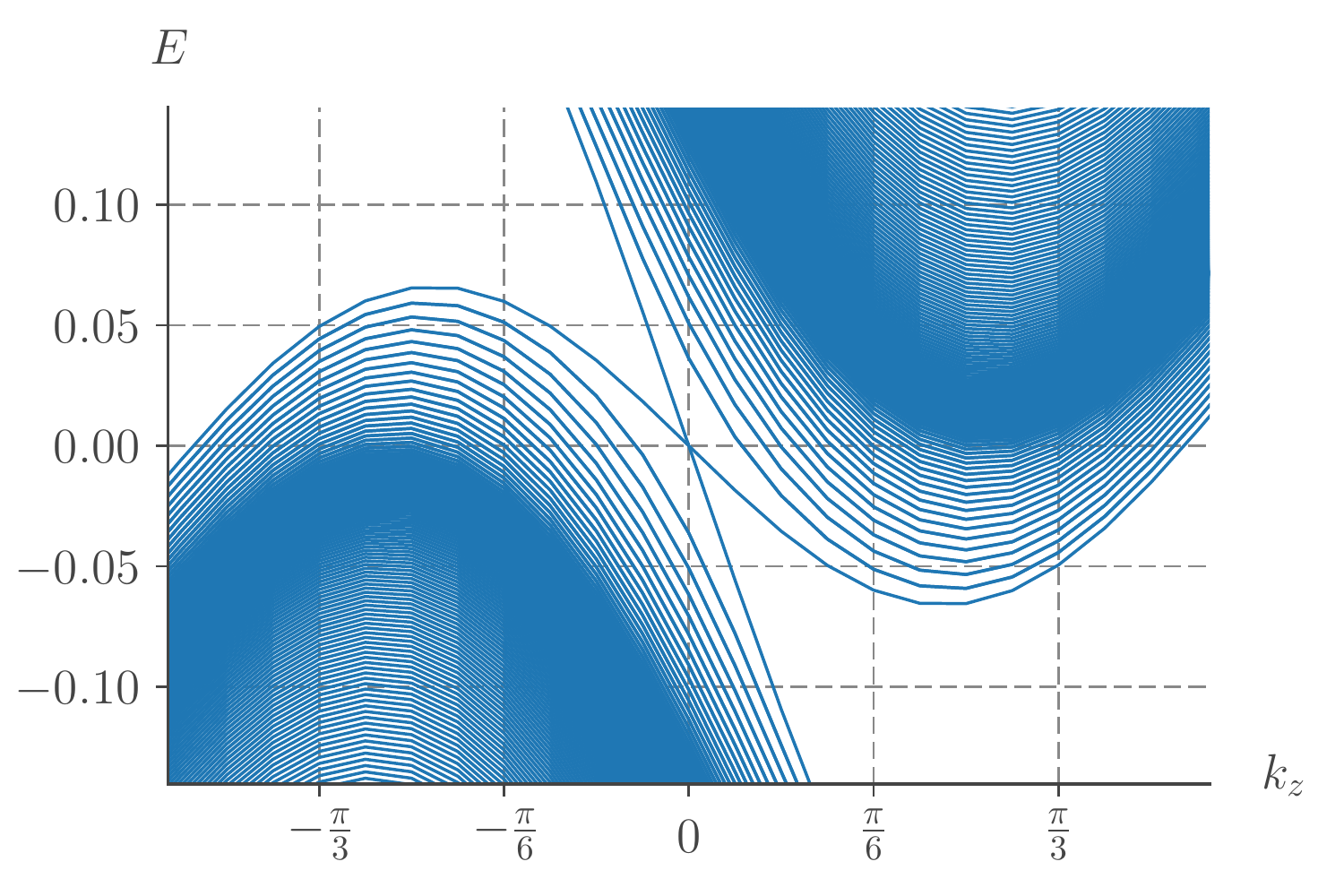}
    \includegraphics[width=0.43\textwidth]{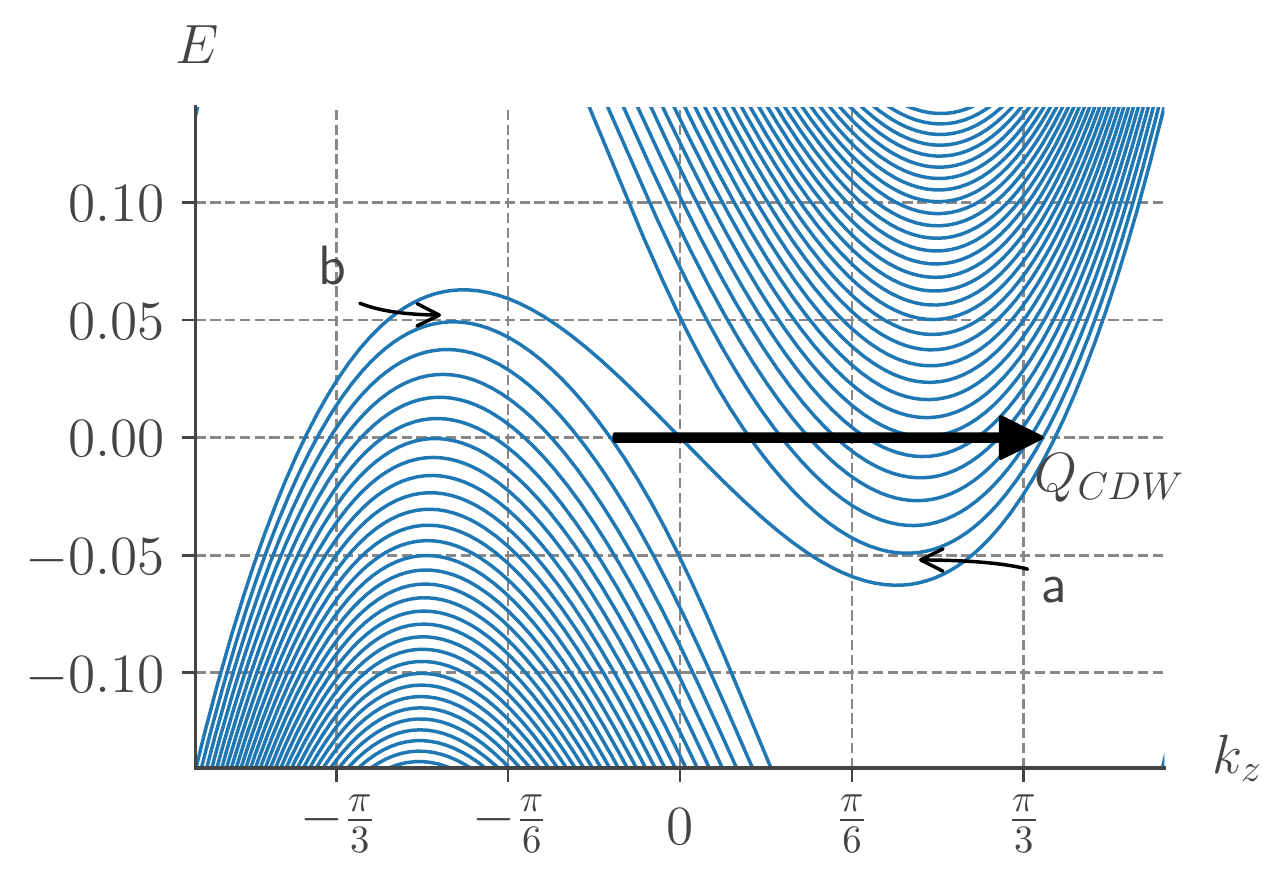}
    \caption{Comparison of the dispersion of the lattice model and the continuum theory without  magnetic field (top panel). The dispersion of the lattice model in a magnetic field is shown in the middle panel (with two Weyl cones at different $k_x$ superposed) and the corresponding dispersion of the continuum theory (single cone) in the bottom panel.
   The parameters of the lattice model are $t_1=-0.8, t_2=-0.6, a_0 = 7 \textrm{\AA}$ and the parameters of the low energy theory are chosen to match the third-order expansion of the lattice model. The magnetic field strength for the bottom row is $B= 86\,T$.}
\label{fig:compare-lattice-continuum}
\end{figure}

Throughout this Rapid Communication we assume the realistic but exemplary values of $\hbar v = 1 \textrm{eV \AA}$ and $a_0 = 7 \textrm{\AA}$.
The term proportional to $t_2$ is included to get smoother electron and hole pockets (see Figure \ref{fig:compare-lattice-continuum}a) and to elucidate how longer range hopping terms can be included in the low order expansion. Our effective model inherits the perfect compensation between electron and hole pockets present in our tight binding model and which is also a prominent feature of WTe$_2$. However, as discussed below such perfect compensation is not a crucial ingredient of our CDW mechanism. 

It is worth mentioning that there are also type-I nodes at $\mathbf k=(\pm \pi/2,0,\pi)$ in this lattice model. However, these are unimportant for our considerations and could easily be removed at the price of a having a more complex lattice model. 

We concentrate on magnetic fields along the $\mathbf{z}$ tilt direction of the cones leading to flat Landau levels in the $k_x, k_y$-plane dispersing only along the $k_z$ direction.
Then we can introduce the field via the usual vector potential $\mathbf{A}$ minimally coupled to the crystal momentum $\mathbf{\Pi}=\hbar \mathbf{k}-\frac{e}{c} \mathbf{A}$ and work with the usual ladder operators in the Landau level (LL) basis [see Supplemental Material (SM)\footnote{See Supplemental Material for details of the mean field calculation and additional information on susceptibility and heat capacity.} for details], such that the spectrum is given by the eigenvalues of 
 
\begin{align}
     \hat H_0^{\text{eff}} \,= &  (-\eta k_z + \gamma k_z^{3}) \sigma_0 + (k_z + \beta k_z^3) \sigma_z + \frac{\sqrt{2 n}}{l_B} \sigma_x
    \label{low-energy}
\end{align}
with the Landau level index $n>0$ and the magnetic length $l_B=\sqrt{\frac{\hbar}{e B}}$. 
The parameters $\eta = -(t_1 + 2t_2),  \beta = -\frac{1}{6}, \gamma = -\frac{1}{6} (t_1 + 8t_2)$ are directly related to our lattice model, with $t_1=-0.8, t_2=-0.6$ for concreteness throughout this paper.

Furthermore, we obtain the transformation relating our original sublattice creation operators $c_{A/B,n,p,k_z}$, with $p$ labeling the degenerate states within each LL, to new operators $a/b_{n,p,k_z}$ for to the electron and holelike bands \begin{align}
    \begin{pmatrix}
        a_{n, p, k_z} \\
        b_{n, p, k_z}
    \end{pmatrix}
    &= 
    \underbrace{
    \begin{pmatrix}
        u(k_z, n, B) & v(k_z, n, B) \\
        -v(k_z, n, B) & u(k_z, n, B)
    \end{pmatrix}}_{\hat U(k_z, n, B)}
    \begin{pmatrix}
        c_{A,n,p,k_z}\\
        c_{B,n,p,k_z}
    \end{pmatrix}
    \label{operator-transformation}
\end{align}
with $u(k_z, n, B)^2 + v(k_z, n, B)^2 = 1$.  Their dispersions are given by
\begin{align}
    E_{a/b}(k_z, n) &= -\eta k_z  + \gamma k_z^3 \pm \sqrt{(k_z + \beta k_z^3)^2 + \frac{2}{l_B^2}|n|}
    \intertext{for $n\neq 0$, and for $n=0$ the chiral level is given by}
    E_{k_z, 0} &= (1 - \eta) k_z + (\beta + \gamma) k_z^3.
\end{align}
The LL dispersions are shown in the bottom panel of Fig.~\ref{fig:compare-lattice-continuum} which can be directly compared to the corresponding numerical tight-binding calculation in a field displayed in the middle panel. Both make apparent one of the main findings of our work --- the size and shape of the two inverted pockets almost exactly match when shifted by the arrow Q$_{CDW}$ indicated in Fig. \ref{fig:compare-lattice-continuum}. These nesting features between entire electron and hole pockets, $E_a(k_z)\approx -E_b(k_z+Q_{CDW})$, are, for example, well established in parent compounds of iron-based superconductors, where they lead to density wave instabilities even for small interactions \cite{Chubukov2008}. This is of course similar to the usual one-dimensional Peierls instability, which is cut off here by the broken inversion symmetry, but arguably more general.

A direct calculation of the corresponding LL- and pocket-resolved susceptibility (see Fig. S1 in the SM) confirms the qualitative picture: (i) A dominant peak at Q$_{CDW}$ appears in the interband component connecting electron and hole pockets; (ii) the peak is maximal for scattering between bands with the same LL index $n$; (iii) due to a small asymmetry between the pockets, which also depends on the LL index, nesting is not perfect which cuts off a true  singularity of the susceptibility and introduces a peculiar field dependence. Note that in the case of imperfect particle-hole compensation the pockets would be shifted in energy with respect to each other leading to dominant nesting between branches with different LL indices, but the overall pictures remains valid.  

{\it Mean Field Ansatz.}
To study the formation of a CDW we add  interactions to the single particle theory presented above.
Due to the multiband nature, even the simplest  contact interaction (strength $U$) is a sufficient approximation for a short-range interaction, as we are interested in the coupling between two different bands. We project the contact interaction to LLs,
\begin{widetext}
\begin{align}
\label{ContInt} 
H_{\textrm{int}}\! =\! 
    \frac{U}{2}  \!
  \sum_{\substack{n_1,n_2,n_3,n_4,\\p_1,p_2,k_z,k_z',\\q_x,q_y,q_z}} \! \!
     e^{iq_y ( p_1 - p_2 - q_x)} 
         J_{n_4,n_1}(\mathbf{q}) J_{n_3,n_2}(-\mathbf{q}) \!\!
    \sum_{\alpha,\beta=A,B} \!
     c^{\dagger}_{\alpha,n_1,p_1,k_z} c^{\dagger}_{\beta,n_2,p_2,k_z'} 
     c_{\beta,n_3,p_2+q_x,k_z'+q_z} c_{\alpha,n_4,p_1-q_x,k_z-q_z} 
 \end{align}
which introduces additional momentum dependence~\cite{Goerbig_RMP} (see SM).

From the main peaks of bare susceptibility we know that the leading CDW instability arises between electron and hole bands with the very same LL index $n$. This allows us to simplify the problem considerably by only considering interactions with all $n_{1,2,3,4} = n$ equal. Hence the different Landau levels  decouple, and we perform the following computations for a fixed Landau level index and combine results for different Landau levels later. Furthermore, since the nesting connects the different branches (with creation operators $a$ and $b$), we are interested in a CDW in $\mean{a^\dagger b}$.
We introduce the generic CDW wavevector $\mathbf{Q} = (Q_x, Q_y, Q_z)$ to formulate the general mean field theory using the ansatz:
$\mean{a^{\dagger}_{p, k_z} b_{p-q_x, k_z-q_z}} = \Delta(k_z, \vec{Q}) e^{-i p Q_y} e^{i Q_y q_x/2} \delta(q_x - Q_x) \delta(q_z - Q_z)$.

Our focus are CDWs along the $k_z$ direction and therefore we concentrate on CDW vectors $\vec{Q} = \begin{pmatrix}
0, 0, Q
\end{pmatrix}$.
We decouple the interaction in the usual way which allows us to write the Hamiltonian in the bilinear form for each LL
\begin{align}
     H_{\textrm{MF}, n}\left( p, k_z \right) &= 
     \begin{pmatrix}
        a^{\dagger}_{p,k_z} & b^{\dagger}_{p,k_z - Q}
    \end{pmatrix}
        \begin{pmatrix}
            E_a(p, k_z) & P(k_z) \\
            P(k_z)^* & E_b(p, k_z-Q)
        \end{pmatrix}
    \begin{pmatrix}
        a_{p,k_z} \\ b_{p,k_z-Q}
    \end{pmatrix}. 
    \label{hmf}
\end{align}
 \end{widetext}
The details of the derivation of the off-diagonal elements $P$ in terms of the projected interaction matrix elements~\cite{goerbig_competition_2004} and the order parameter $\Delta$ are given in the SM. There, we also show that our ansatz gives real electron densities~\cite{fukuyama_two-dimensional_1979}. Knowing $P$, we obtain self-consistent solutions for $\Delta$ numerically and thereby determine whether or not the system supports a CDW. 

As a check, we have confirmed that the wave vector corresponding to the smallest critical interaction for a mean-field CDW transition indeed coincides with the main peak of the bare susceptibility at $Q=Q_{CDW}$.

{\it Results: Cascade of CDW transitions in temperature.}
\begin{figure}[tb]
    \centering
    \includegraphics[width=0.35\textwidth]{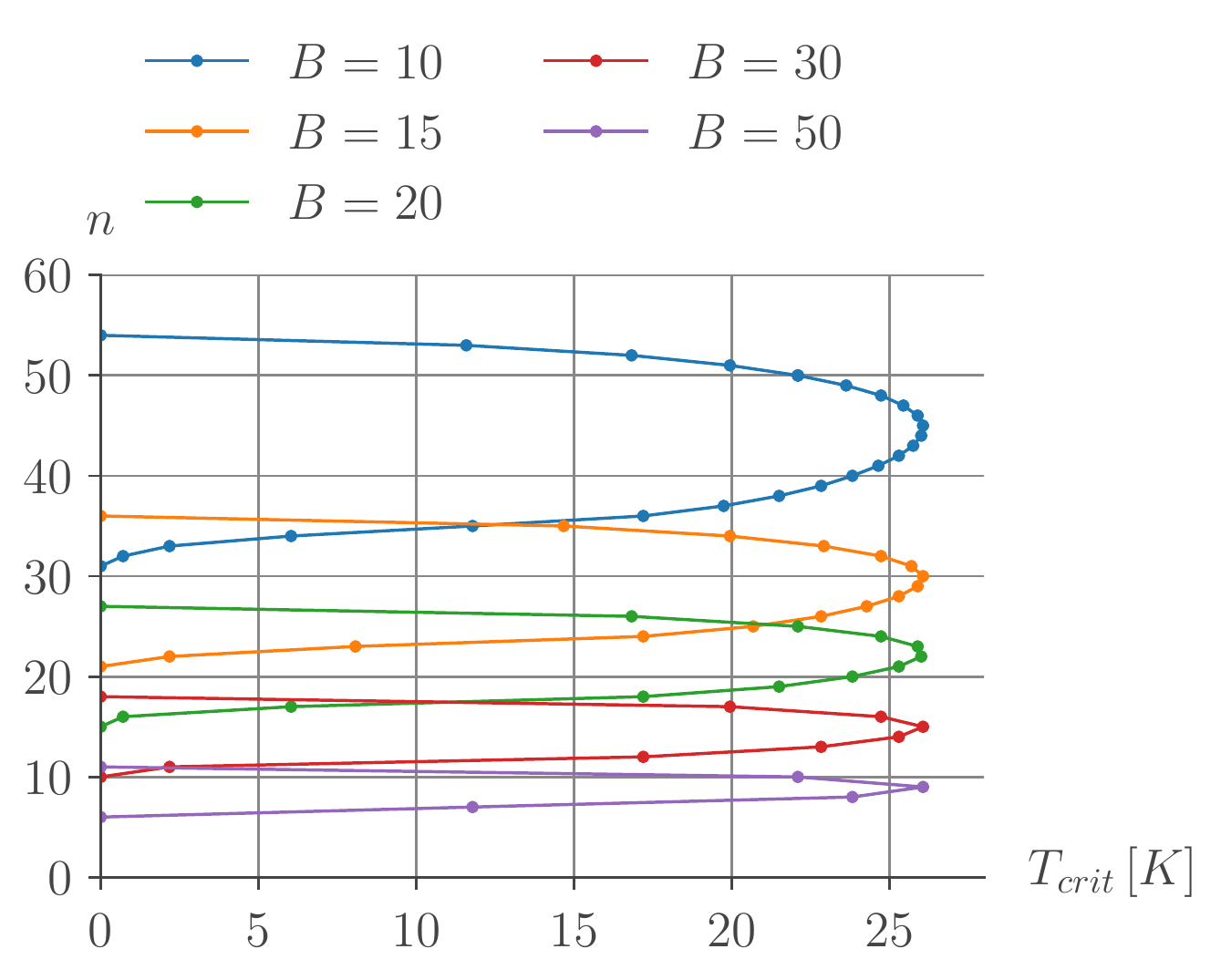}
    \includegraphics[width=0.35\textwidth]{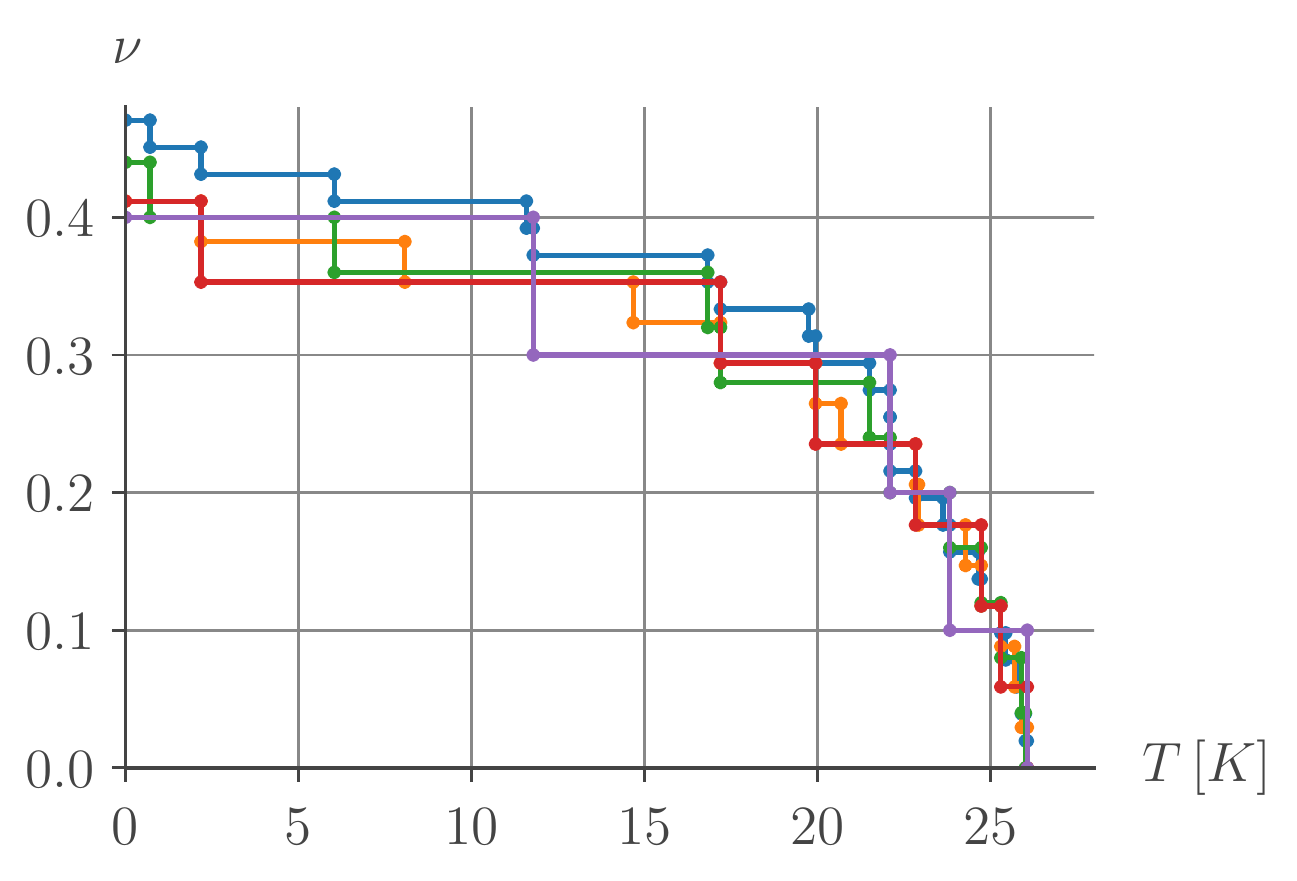}
    \caption{Top panel: Highest critical temperature per Landau level for fixed interaction strength $U=0.025$ and fixed wave vector $Q_{CDW}=1.48$ (see discussion in the main text).
    Bottom panel: Fraction of levels crossing the Fermi energy (when no interaction is considered) that are gapped by a CDW. When lowering the temperature, there is a cascade of consecutive CDW transitions, leading to an increased fraction of gapped levels. Parameters are the same as for the top panel.}
    \label{fig:tcrit}
\end{figure}
While an independent determination of the CDW wave vector ($Q_{CDW}$) for each Landau level is possible we choose a common $Q_{CDW}$ for all Landau levels to account for the inevitable inter-Landau-level-coupling, which we neglected in our approximation.
This global $Q_{CDW}$ is obtained by maximizing the number of gapped levels, and hence represents the energetically most favorable configuration.

In Fig. \ref{fig:tcrit} we show the numerical results for the critical temperature of the CDW transition at different magnetic fields and per Landau level $n$.
It becomes clear that in lowering the temperature more and more, the Landau levels undergo the phase transition, hence we observe a {\it cascade of successive CDW transitions}.
It is important to keep in mind that the magnetic field also changes the spacing between Landau levels and thereby the number of Landau levels crossing the Fermi energy (the degeneracy in each level increases accordingly).
To account for this, we count the total number of gapped levels at each temperature and magnetic field strength and compute the fraction of this number compared to the number of Landau levels crossing the Fermi energy in the corresponding noninteracting system. This fraction is shown in the bottom panel of Fig. \ref{fig:tcrit} and turns out to be roughly independent of  field.

\begin{figure}[ht]
    \centering
    \includegraphics[width=0.35\textwidth]{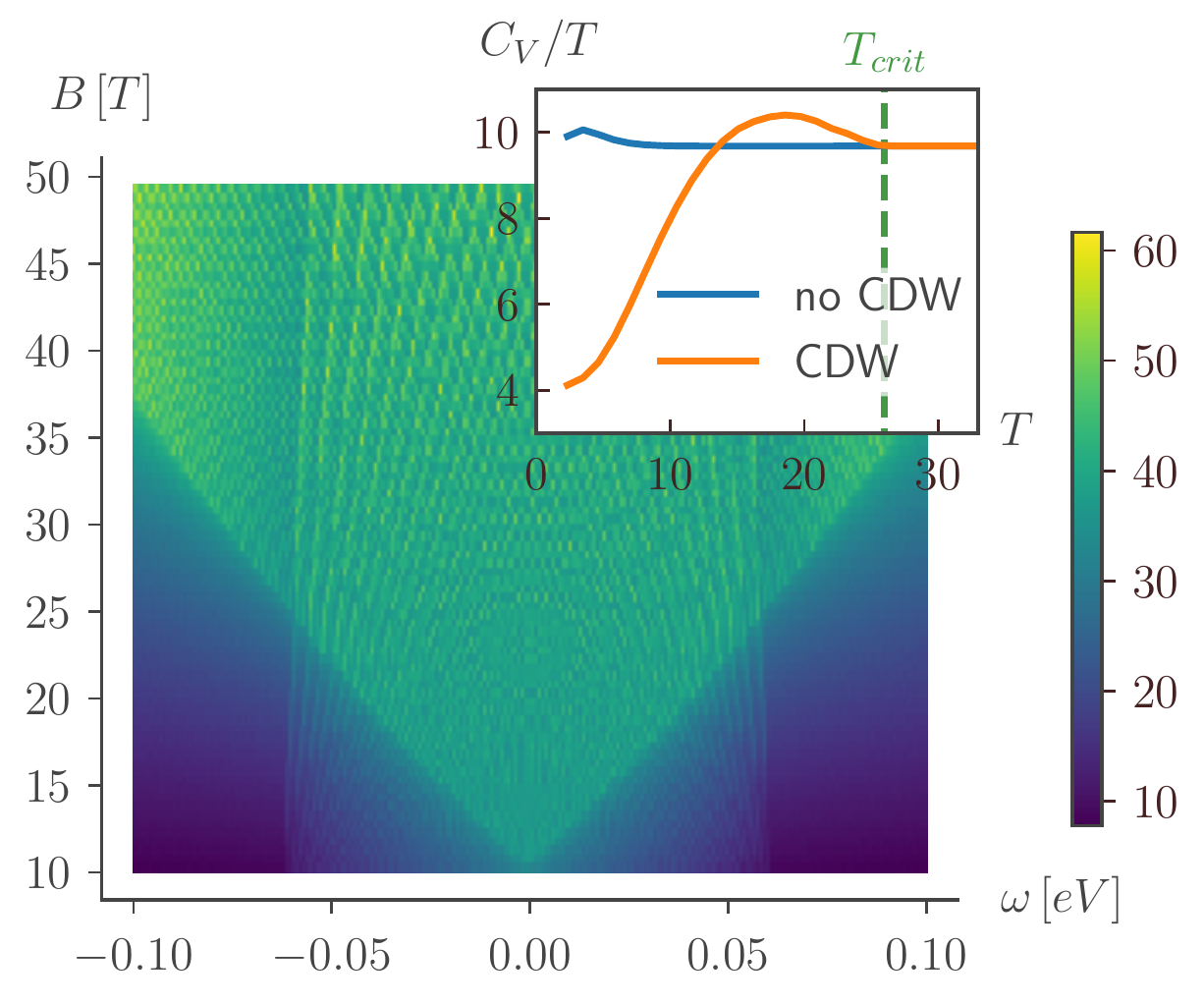}
    \includegraphics[width=0.35\textwidth]{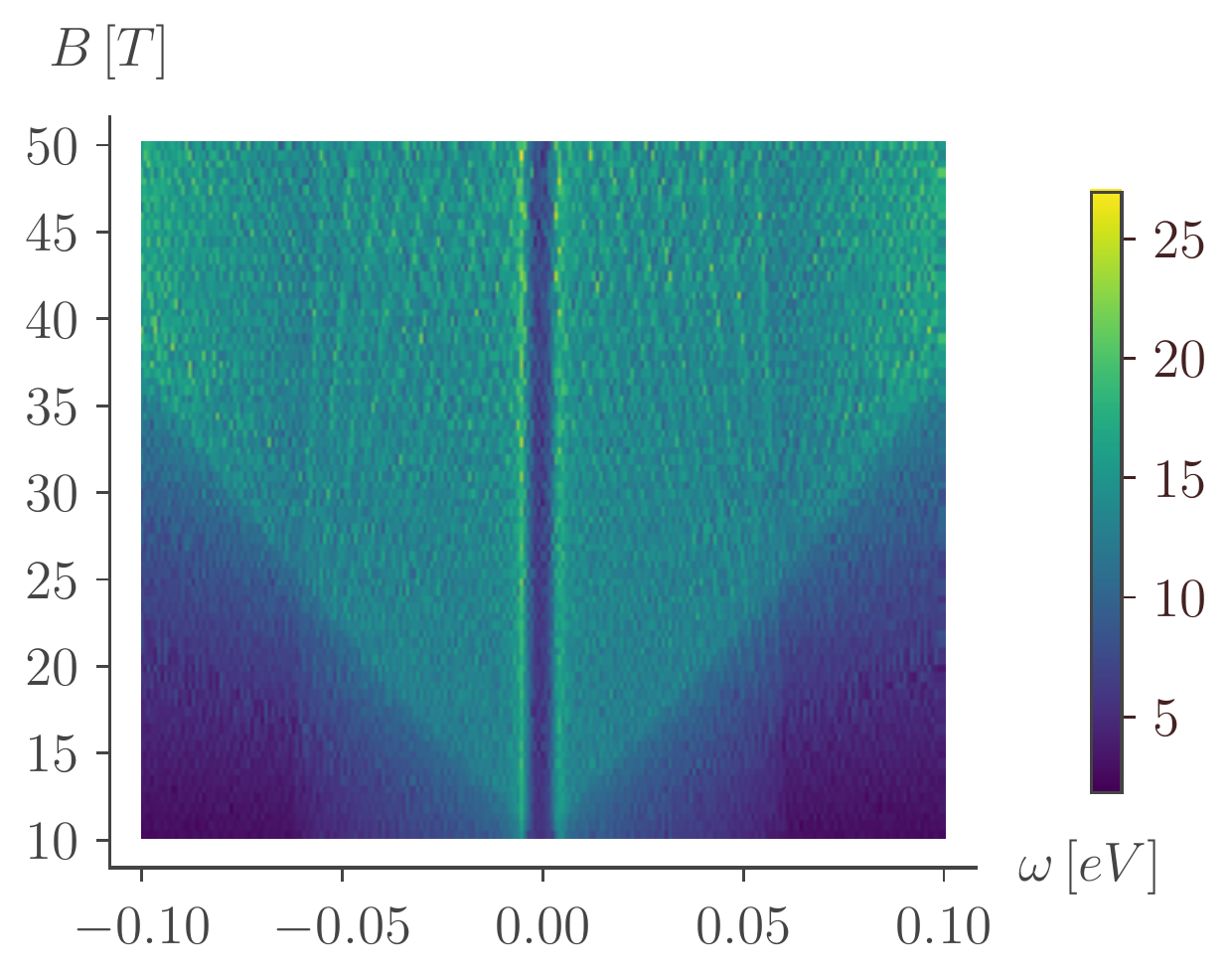}
    \caption{DOS as a function of of magnetic field strength and energy, summed over the 50 lowest Landau levels. When including interactions (bottom panel, numerical result obtained at $T=4.6K$) a gap opens compared to the case without interactions (top panel). Inset: Temperature dependence of heat capacity $C_V/T$ (arbitrary units) at fixed magnetic field of $B=20 T$. As expected $C_V/T$ is constant (blue line) without the CDW, but taking the CDW into account (orange line) deviations are clearly visible below the critical temperature indicated by the dashed green line.}
    \label{fig:dos}
\end{figure}

{\it Experimental signatures.}
Our scenario implies thermodynamic signatures of a high-field phase transition, e.g., in specific heat or magnetization (see the inset in Fig. \ref{fig:dos}). The CDW real-space modulation of the electronic density $\rho (\mathbf{r}) \propto \cos (Q_{CDW}\cdot r_z)$ should be detectable via x-ray scattering.   

In addition, the suppression of electronic states around the chemical potential entails clear experimental signatures and should be observable, for example, via scanning tunneling microscopy (STM) measurements. In Fig. \ref{fig:dos} we compare the energy-resolved density of states (DOS) as a function of  magnetic field between a noninteraction system (top panel) and the system with weak interactions (bottom panel). We observe a striking difference at zero energy (near the Weyl point) where the CDW clearly leads to a strong suppression of the DOS. As not all Landau levels are gapped by the CDW, for small interactions there is some residual DOS in this region, and hence it is not a complete gap. 

Note that the DOS oscillates due to the discreteness of Landau levels. We expect that the corresponding quantum oscillations in thermodynamic observables survive even in regimes in which the CDW opens a full gap~\cite{Knolle_2015}.

Finally, if the magnetic field is not aligned in the direction of the tilt, as chosen in our setup, the electron- and holelike pockets of the LLs disappear \cite{Udagawa2017}, leading to a characteristic suppression of the CDW.

{\it Discussion.} We introduced an exemplary lattice model of a type-II Weyl semimetal and obtained a low energy description that takes finite electron and hole pockets into account. From this we identified emergent nesting properties that occur between electron- and holelike Landau level branches once the Weyl semimetal is placed in a magnetic field. We developed a general mean-field theory of a Weyl semimetal in a field, confirming the intuitive picture of a nesting induced CDW instability. The self-consistent calculations allowed us to trace its dependence on temperature and magnetic field.

Here, we focused on the most relevant parts, i.e., intra Landau level couplings between different branches. While we made plausible why these approximations should be valid, a more quantitative analysis of the inter-Landau-level couplings poses interesting questions for future research, as well as a full lattice calculation. 
We have pointed out several clear experimental signatures of the field-induced CDW transition, e.g., in thermodynamics, STM, and xrays.

The reduction of the DOS would also lead to an increased magnetoresistance, which is a feature of great interest in many type-II Weyl systems. 
A similar mechanism has been proposed to explain the magnetoresistance properties of graphite~\cite{yoshioka_electronic_1981,fauque_two_2013,leboeuf_thermodynamic_2017}.
Considering the parallel alignment of the tilt and the magnetic field, our scenario is only directly applicable to WP$_2$ \cite{Autes2016,Kumar2017}, while the geometry is different in the case of WTe$_2$ \cite{ali_large_magnetoresistance_wte2_2014} and  Mo$_x$W$_{1-x}$Te$_2$ \cite{belopolski_discovery_2016}. A more detailed study of transport properties is desirable and left for future research.
However, if our scenario is \textit{mutatis mutandis} applicable to the nonsaturating magnetoresistance of these materials as well, there are some immediate consequences: Most saliently it would be in contrast to the semiclassical picture suggested in Ref. \onlinecite{ali_large_magnetoresistance_wte2_2014} which relies on on strict particle-hole symmetry. It is supported by the fact that the magnetotransport of WTe$_2$ has an unusual temperature dependence and that MoTe$_2$ with similar properties is far from particle-hole compensated~\cite{DDSarma_2017}. Note that the number of LLs crossing the Fermi level is determined by the magnetic field component $B^\perp$ projected along the tilt direction. Hence, a magnetoresistance from LL formation of the form $(B^\perp)^2$ suggests a $\cos^2 \theta$ angular dependence similar to measurements on WP$_2$ \cite{Kumar2017}.

Finally, we note that for larger interactions the LL spectrum is fully gapped, which should lead to a concomitant three-dimensional Hall plateau \cite{Bernevig2007} similar to other systems with density-wave-induced three-dimensional Hall effects~\cite{McKernan1995,Balicas1995}.

\begin{acknowledgments}
This work was supported by Emmy
Noether program (BE 5233/1-1) of the Deutsche Forschungs-
gemeinschaft, the Swedish Research Council (VR), and
the Wallenberg Academy Fellows program of the Knut
and Alice Wallenberg Foundation. J.K. is supported by the
Marie Curie Programme under EC Grant Agreement No.
703697.
\end{acknowledgments}

\onecolumngrid
\renewcommand\thefigure{S\arabic{figure}}    
\setcounter{figure}{0} 
\section{Supplementary Material for ``Charge density wave instabilities of type-II Weyl semimetals in a strong magnetic field''}
\section{Susceptibility predictions for $Q_{CDW}$}
\label{app:susceptibility}
\begin{figure}[ht]
	{
      \centering
      \includegraphics[height=0.25\textwidth]{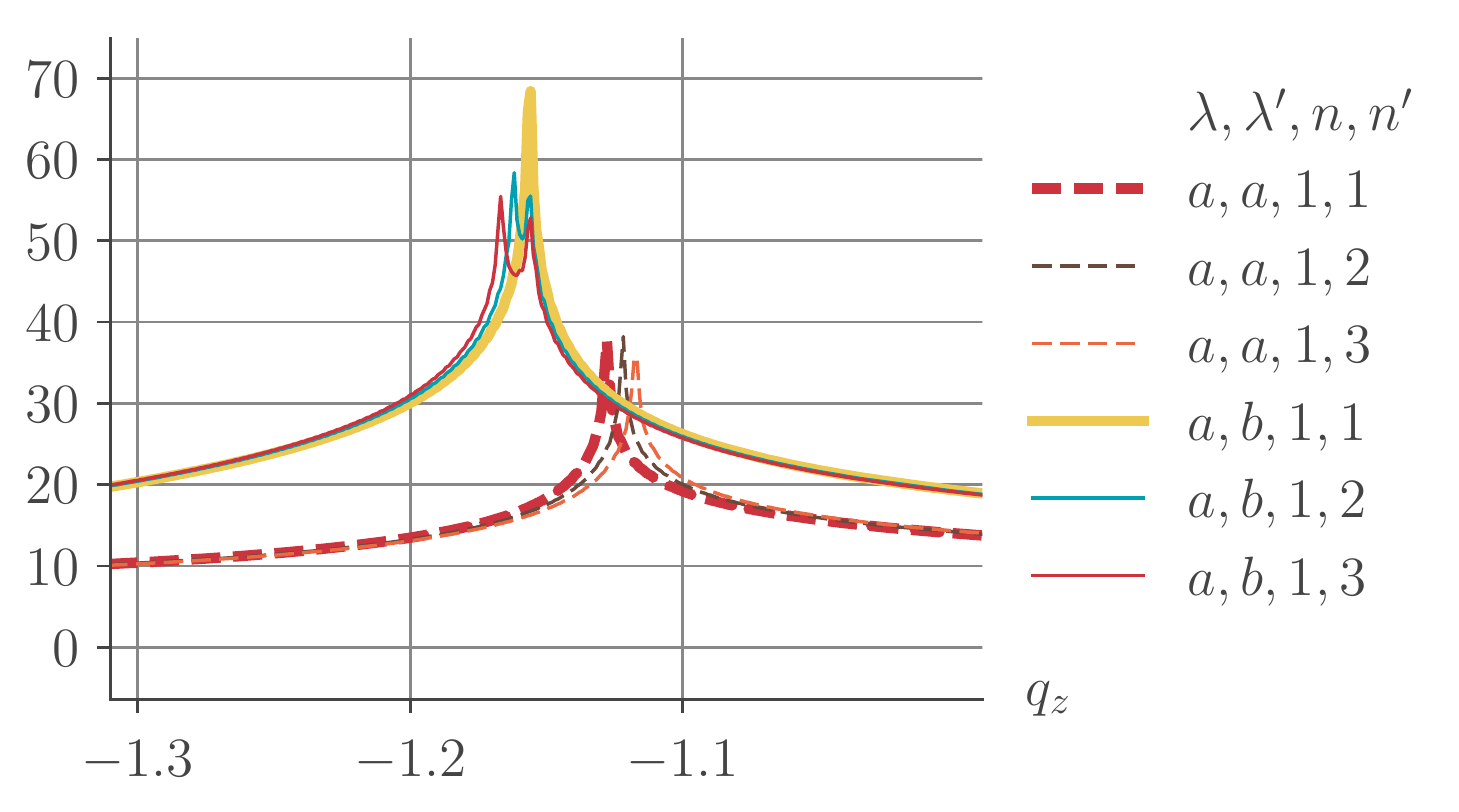}
      \includegraphics[height=0.25\textwidth]{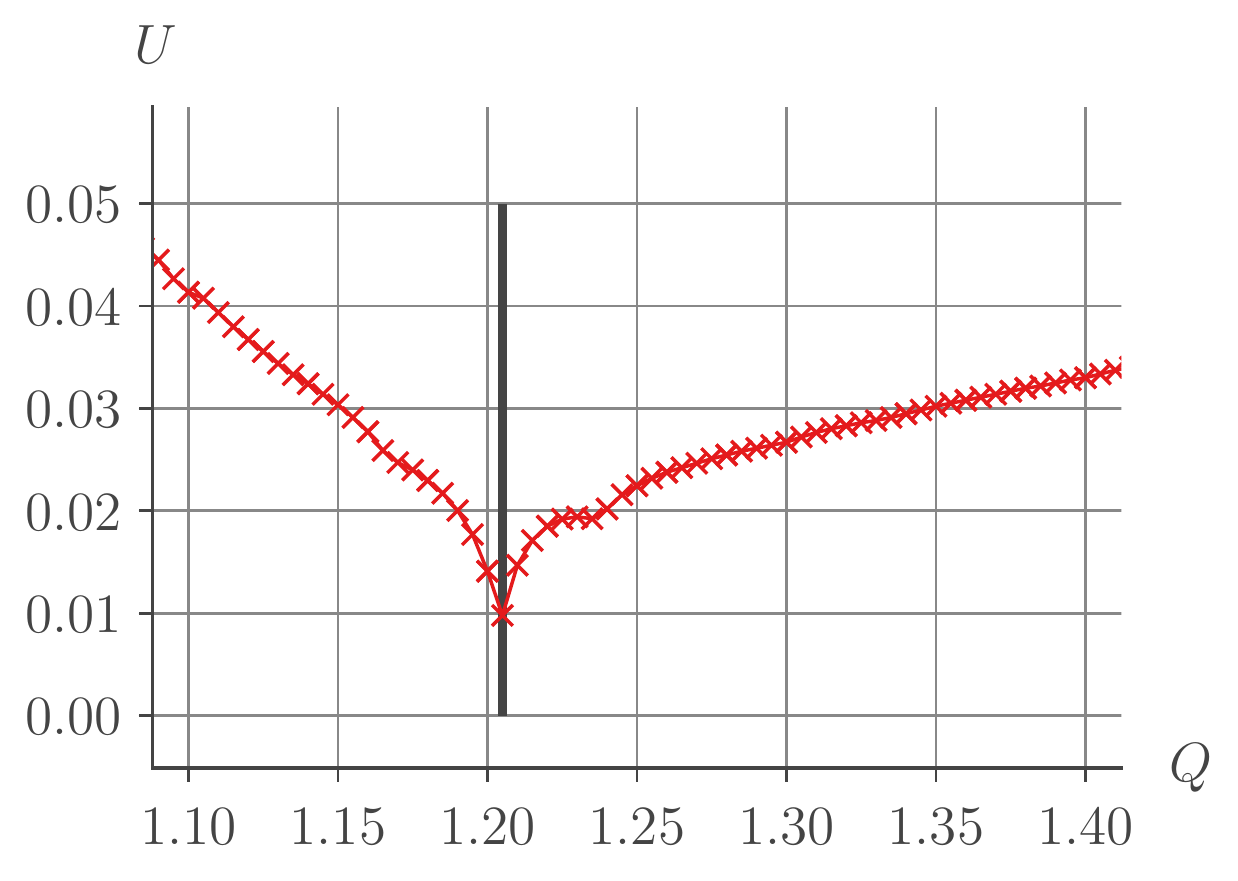}
    }
    \caption{Left panel: Susceptibility (arbitrary units) in the non-interacting case with parameters at $B=20 \textrm{T}$, $T=0.46 \textrm{K}$, $\omega=0$.
        The maximal susceptibility is reached for $\lambda = a, \lambda' = b$ and $n=n'$.
    Right panel: Critical interaction strength in dependence of CDW wavevector $Q$ for a single Landau level ($n=1, B=10$). The solid vertical line corresponds to the prediction from the susceptibility calculation.
    }
    \label{fig:susceptibility}
\end{figure}

We use the standard formula to compute the susceptibility\cite{goerbig_competition_2004}
\begin{align}
    \Pi_{nn'}^{\lambda \lambda'}\left( \vec{q}, \omega \right) = \lim_{\delta\rightarrow 0} \sum_{\vec{q}'} \frac{n_F\left(E_{\lambda}(\vec{q}', n)\right) - n_F\left( E_{\lambda'}\left( \vec{q}' + \vec{q}, n' \right) \right)}{E_{\lambda}(\vec{q}', n) - E_{\lambda'}(\vec{q}'+\vec{q}, n') + \omega + i \delta}
    \label{}
\end{align}
using the Fermi function $n_F$, which depends on temperature, $\Pi$ is shown for some representative parameters in Fig. \ref{fig:susceptibility}. In the right panel of this figure we show that $q$ corresponding to the highest peak in susceptibility is indeed the optimal $Q_{CDW}$ (lowest critical interaction strength) for the charge density wave.

\section{Landau level spectrum and interaction}
\label{app:reference}
As discussed in the main text, the magnetic field $\vec{B} = \nabla \times \vec{A}$ is introduced by a minimal coupling of the vector potential, which we consider to be $\vec{A} = \left( 0, Bx, 0 \right)$ in the Landau gauge, to the crystal momentum $\vec{\Pi} = \hbar \vec{k} - \frac{e}{c}\vec{A}$.
Then we obtain the standard raising and lowering operators $a = \frac{l_B}{\sqrt{2} \hbar} \left( \Pi_x - i \Pi_y \right)$ and $a^\dagger = \frac{l_B}{\sqrt{2}\hbar} \left(\Pi_x + i \Pi_y\right)$ with $[a, a^\dagger] = 1$. And consequently we arrive at the following Hamiltonian describing the low energy theory
\begin{align}
    H_0 = \int \d^3 r 
    \begin{pmatrix}
        \Psi_A^\dagger(\vec{r})
         & \Psi_B^\dagger(\vec{r})
    \end{pmatrix}
    \begin{pmatrix}
        (1-\eta)k_z + (\gamma + \beta) k_z^3 & \frac{\sqrt{2}}{l_B} \hat{a}^\dagger \\
        \frac{\sqrt{2}}{l_B} \hat{a} & -(1+\eta)k_z + (\gamma - \beta) k_z^3
    \end{pmatrix}
    \begin{pmatrix}
        \Psi_A(\vec{r}) \\
         \Psi_B(\vec{r})
    \end{pmatrix}
    \label{}
\end{align}
with
\begin{align}
    \Psi_A^\dagger(\vec{r}) &= \sum_{n,p,k_z} e^{i k_z z} \psi_{n,p}(x,y) \hat{c}_{A, n, p, k_z}\\
    \psi_{n,p}(x,y) &= \frac{1}{\sqrt{L}} e^{i p x } \left( \pi 2^{2n} (n!)^2 \right)^{-1/4} e^{-\frac{1}{2} (y+p)^2} H_n(y+p)
    \intertext{while for $B$ there is a shift from $n$ to $n-1$}
    \Psi_B^\dagger(\vec{r}) &= \sum_{n,p,k_z} e^{i k_z z} \psi_{n-1,p}(x,y) \hat{c}_{B, n, p, k_z}
    \label{wavefunctions}
\end{align}
where $\psi$ are properly normalized Harmonic oscillator wavefunctions including the Hermite polynomials $H_n$.

The generic interaction is given by
\begin{align}
H_{\textrm{int}} &= \frac{1}{2}  \sum_{\substack{\alpha,\beta=A,B}} \int \d{^3\vec{r}_1} \int \d{^3 \vec{r}_2} \Psi^\dagger_\alpha(\vec{r}_1) \Psi^\dagger_\beta(\vec{r}_2) U( \vec{r}_1 - \vec{r}_2) \Psi_\beta(\vec{r}_2) \Psi_\alpha(\vec{r}_1)
\end{align}
where the contact interaction corresponds to the interaction potential
$U(\vec{r}) = \delta(\vec{r})$. Note, in using the contact interaction we avoid introducing another scale, but we checked that the qualitative findings presented in the following can be reproduced using longer ranged interactions including Coulomb.

Projecting this contact interaction onto the Landau levels we obtain Eq. 7 in the main text with
\begin{align}
\label{ContIntAppendix}
    H_{\textrm{int}} &= 
    \frac{1}{2} \, 
     \sum_{\substack{n_1,n_2,n_3,n_4,\\p_1,p_2,k_z,k_z',\\q_x,q_y,q_z}}
     e^{iq_y ( p_1 - p_2 - q_x)} M_{n_1,n_2,n_3,n_4}(\vec{q})
     \sum_{\alpha,\beta=A,B} 
     c^{\dagger}_{\alpha,n_1,p_1,k_z} c^{\dagger}_{\beta,n_2,p_2,k_z'} 
     c_{\beta,n_3,p_2+q_x,k_z'+q_z} c_{\alpha,n_4,p_1-q_x,k_z-q_z} 
 \end{align}
 \begin{align}
    M_{n_1,n_2,n_3,n_4}(\vec{q}) &=
        U(\vec{q})
        \underbrace{ J_{n_4,n_1}(q_x, q_y) J_{n_3,n_2}(-q_x, -q_y) }_{F(q_x, q_y)} 
\end{align}
with $c_{\alpha=B, n} \rightarrow c_{B,n+1}$, the Fourier transformed interaction potential $U(\vec{q})$ and the factor $F$ due to the wavefunctions overlap. Here all in-plane momenta are measured in units of $l_B$.
Using the wavefunctions from  Eq. \eqref{wavefunctions} we get for $J$~\cite{bychkov_two-dimensional_1983}
\begin{align}
    J_{m,n}(q_x, q_y) &= \sqrt{\frac{n!}{m!}} e^{-|q|^2/4} \left( \frac{q_x + i q_y}{\sqrt{2}} \right)^{m-n} L^{m-n}_n \left( \frac{|q|^2}{2}\right)  \quad \textrm{with } |q|^2 = q_x^2 + q_y^2 \quad \textrm{for } \; m < n \\
    J_{m,n}\left(q_x, q_y\right) &= J_{n,m}^{*}\left( -q_x, -q_y \right) .
\end{align}

The full Hamiltonian then reads
\begin{align}
H = H_0 + H_{int} \,. \label{hamiltonian}
\end{align}

\section{Density wave order parameter}
\label{app:derivations}
\label{app:ansatz}
The intra LL branch components
\begin{align}
    \mean{a^{\dagger}_{p, k_z} a_{p-q_x, k_z-q_z}} &= n_a(k_z) e^{-i p Q_y} e^{i Q_y q_x/2} \delta(q_x-Q_x)\delta(q_z) \\
    \mean{b^{\dagger}_{p, k_z} b_{p-q_x, k_z-q_z}} &= n_b(k_z) e^{-i p Q_y} e^{i Q_y q_x/2} \delta(q_x - Q_x)\delta(q_z)
\end{align}
only shift the chemical potential which can be fixed independently.
To motivate the ansatz for our CDW order parameter, we show in the following explicitly that the resulting electron density is real and modulated in the form $\cos(Q_z\cdot r_z)$.

For now we look only at the density of the $c_A$ part and we fix $n$ and $B$ so we do not write out dependencies on these variables:
\begin{align}
    \rho_A(\vec{r}) &= \sum_{\vec{q}}{e^{i \vec{q} \cdot \vec{r}} \rho_A(\vec{q})}
    \intertext{where}
    \rho_A(\vec{q}) &= \sum_{p_1, k_z}{ e^{i q_y (p_1 - q_x/2)} \mean{c^{\dagger}_{A,p_1,k_z}c_{A,p_1-q_x,k_z-q_z}}} 
    \intertext{so that}
    \rho_A(\vec{r}) &= \sum_{\vec{q}}{e^{i \vec{q} \cdot \vec{r}} \sum_{p_1, k_z}{ e^{i q_y (p_1 - q_x/2)} \mean{c^{\dagger}_{A,p_1,k_z}c_{A,p_1-q_x,k_z-q_z}}}} \\
   &= \sum_{\vec{q}}{e^{i \vec{q} \cdot \vec{r}} \sum_{p_1, k_z}{ e^{i q_y (p_1 - q_x/2)} \mean{(u(k_z) a^{\dagger}_{p_1,k_z} - v(k_z) b^{\dagger}_{p_1,k_z})(u(k_z-q_z) a_{p_1-q_x,k_z-q_z} - v(k_z-q_z)b_{p_1-q_x,k_z-q_z}}}} \\
       &= \sum_{\vec{q}}{e^{i \vec{q} \cdot \vec{r}} \sum_{p_1, k_z}{ e^{i q_y (p_1 - q_x/2)} }}
   \Big(
   u(k_z) u(k_z-q_z) \mean{a^{\dagger}_{p_1,k_z} a_{p_1-q_x,k_z-q_z}} \label{part-aa}\\
   & + v(k_z) v(k_z-q_z) \mean{b^{\dagger}_{p_1,k_z} b_{p_1-q_x,k_z-q_z}} \label{part-bb}\\
   & + u(k_z) v(k_z-q_z) \mean{a^{\dagger}_{p_1,k_z} b_{p_1-q_x,k_z-q_z}} \label{part-ab}\\
   & + v(k_z) u(k_z-q_z) \mean{b^{\dagger}_{p_1,k_z} a_{p_1-q_x,k_z-q_z}}
   \Big) 
    \label{part-ba}
\end{align}

where the first two summands, Eqs. \eqref{part-aa} and \eqref{part-bb}, are just given by $\sum_{k_z} u(k_z)^2 n_A(k_z) + v(k_z)^2 n_B(k_z)$.
So we consider the second part in detail, Eqs. \eqref{part-ab} and \eqref{part-ba}, using the ansatz from above. We obtain the corresponding $\mean{b^\dagger a}$ terms by shifting variables:

\begin{align}
    \sum_{\vec{q}}{e^{i \vec{q} \cdot \vec{r}} \sum_{p_1, k_z}{ e^{i q_y (p_1 - q_x/2)} }} \Big(
    & u(k_z) v(k_z-q_z) \Delta(k_z, \vec{Q}) e^{-i p_1 Q_y} e^{i Q_y q_x/2} \delta(q_x - Q_x) \delta(q_z - Q_z) \\
    &+ v(k_z) u(k_z-q_z) \Delta^{*}(k_z+q_z, \vec{Q}) e^{i p_1 Q_y} e^{-i Q_y q_x/2} \delta(q_x + Q_x) \delta(q_z + Q_z)\Big)
    \intertext{then $\sum_{p_1} e^{i p_1 (q_y \pm Q_y)}$ gives a $\delta(q_y \pm Q_y)$, such that}
    = \sum_{\vec{q}}{e^{i \vec{q} \cdot \vec{r}} \sum_{k_z}{ e^{- i q_y  q_x/2} }} \Big(
    & u(k_z) v(k_z-q_z) \Delta(k_z, \vec{Q}) e^{i Q_y q_x/2} \delta(q_x - Q_x) \delta(q_y - Q_y) \delta(q_z - Q_z) \\
    &+ v(k_z) u(k_z-q_z) \Delta^{*}(k_z-q_z, \vec{Q}) e^{-i Q_y q_x/2} \delta(q_x + Q_x) \delta(q_y + Q_y)\delta(q_z + Q_z)\Big),
    \intertext{making use of all the delta functions and using $\vec{Q} = (Q_x, Q_y, Q_z)$ we get}
    =  \sum_{k_z}\Big(
    & e^{i \vec{Q} \cdot \vec{r}} u(k_z) v(k_z-Q_z) \Delta(k_z, \vec{Q}) \\
    &+e^{-i \vec{Q} \cdot \vec{r}} v(k_z) u(k_z+Q_z) \Delta^{*}(k_z+Q_z, \vec{Q})\Big).
    \intertext{Now we shift the $k_z$ of the second summand to $k_z \rightarrow k_z-Q_z$, as the summation over $k_z$ goes from $-\infty$ to $+\infty$ and $\lim_{k_z \rightarrow \pm \infty} v(k_z) = 0$ this poses no problems. Finally we get}
    \rho_A (\mathbf{r})&=  \sum_{k_z} u(k_z) v(k_z-Q_z) \Big( e^{i \vec{Q} \cdot \vec{r}} \Delta(k_z, \vec{Q}) + e^{-i \vec{Q} \cdot \vec{r}} \Delta^{*}(k_z, \vec{Q})\Big)
    \label{}
\end{align}
which is manifestly real.

\subsection{Fock term}
\label{app:fock}
In the mean field expansion we need to pay special attention to the Fock terms and we use a trick from Ref. \cite{goerbig_competition_2004}. Starting from an exemplary Fock term in the mean field expansion
\begin{align}
    &\sum_{\vec{q}, p_1, p_2, k_z, k_z'} e^{i q_y (p_1 - p_2 - q_x)} M(\vec{q}) \mean{c^{\dagger}_{p_1, k_z} c_{p_2+q_x, k_z'-q_z}} c^{\dagger}_{p_2, k_z'} c_{p_1-q_x, k_z-q_z}
\end{align}
following the steps of \cite{goerbig_competition_2004} we arrive at
\begin{align}
    =& \quad \sum_{p_x, p_y, s_z, } \frac{1}{N_B} \sum_{q_x, q_y} e^{i (q_y p_x - p_y q_x)} M(\vec{q}) \sum_{y_+, y_-} e^{i p_y (y_+ - y_- - p_x)} \mean{c^{\dagger}_{y_+, k_z} c _{y_+ - p_x, k_z - s_z}} c^{\dagger}_{y_-, k_z'} c_{y_-+p_x, k_z'+s_z} 
\end{align}
where we used the replacements $p_x = p_1 - p_2 - q_x, 
\quad R= \frac{p_1+p_2}{2}, 
        \quad y_\pm = R \pm \left(  \frac{q_x + p_x}{2} \right)$ and inserting the clever 1, $N_B$ being the number of state in the Landau level: 
        $\frac{1}{N_B} \sum_{y_+,p_y} e^{i p_y (y_+ - y_- - q_x - p_x)} = 1$.
(The indices $A,B$ of the c operators are not specified here, but the order of the operators stays fixed).
        
Renaming variables as $y_+ \rightarrow p_1, y_- \rightarrow p_2, s_z \rightarrow q_z, p_x \leftrightarrow q_x, p_y \leftrightarrow q_y$ we get
\begin{align}
    &\sum_{\vec{q}, p_1, p_2, k_z, k_z'} e^{i q_y (p_1 - p_2 - q_x)} \tilde{M}(\vec{q}) \mean{c^{\dagger}_{p_1, k_z} c_{p_1-q_x, k_z-q_z}} c^{\dagger}_{p_2, k_z'} c_{p_2+q_x, k_z'+q_z},
    \label{fock1}
\end{align}
where
\begin{align}
    \tilde{M}(\vec{q}) &= \frac{1}{N_B} \sum_{p_x, p_y} e^{i(p_y q_x - p_x q_y)} M(\vec{p})
    \label{fock2}
\end{align}
which is (up to a rotation) the Fourier transform of $M(\vec{q})$ in the $q_x, q_y$ plane.

\subsection{Projected mean-field matrix elements}
\label{app:matrixelements}

We start with the interaction matrix from above for the Hartree terms, and use \eqref{fock2} for the Fock terms to get the matrix elements in the c-basis.
First we write the Hamiltonian \eqref{hamiltonian} as a matrix:
\begin{align}
    H &= 
    \begin{pmatrix}
        c^\dagger_{A,p_1,k_z} c^\dagger_{A,p_2,k_z'} &
        c^\dagger_{A,p_1,k_z} c^\dagger_{B,p_2,k_z'} &
        c^\dagger_{B,p_1,k_z} c^\dagger_{A,p_2,k_z'} &
        c^\dagger_{B,p_1,k_z} c^\dagger_{B,p_2,k_z'}
    \end{pmatrix}
    H_{c}
    \begin{pmatrix}
        c_{A,p_2+q_x,k_z'+q_z} c_{A,p_1-q_x,k_z-q_z} \\
        c_{A,p_2+q_x,k_z'+q_z} c_{B,p_1-q_x,k_z-q_z} \\
        c_{B,p_2+q_x,k_z'+q_z} c_{A,p_1-q_x,k_z-q_z} \\
        c_{B,p_2+q_x,k_z'+q_z} c_{B,p_1-q_x,k_z-q_z}
    \end{pmatrix}
    \\
    &=
    \begin{pmatrix}
        a^\dagger_{p_1,k_z} a^\dagger_{p_2,k_z'} &
        a^\dagger_{p_1,k_z} b^\dagger_{p_2,k_z'} &
        b^\dagger_{p_1,k_z} a^\dagger_{p_2,k_z'} &
        b^\dagger_{p_1,k_z} b^\dagger_{p_2,k_z'}
    \end{pmatrix}
    H_{a}
    \begin{pmatrix}
        a_{p_2+q_x,k_z'+q_z} a_{p_1-q_x,k_z-q_z} \\
        a_{p_2+q_x,k_z'+q_z} b_{p_1-q_x,k_z-q_z} \\
        b_{p_2+q_x,k_z'+q_z} a_{p_1-q_x,k_z-q_z} \\
        b_{p_2+q_x,k_z'+q_z} b_{p_1-q_x,k_z-q_z}
    \end{pmatrix}
    \intertext{with}
    H_{a} &= U_4(k_z, k_z') \, H_{c} \, U_4^T(k_z'+q_z, k_z-q_z) \\
    \intertext{where}
    U_4(k_1, k_2) &= U(k_1, B) \otimes U(k_2, B)
    \label{}
\end{align}

To obtain the matrix elements $P$ used in the mean-field Hamiltonian in the main text, we write each of the 16 terms of this Hamiltonian in a mean field expansion and collect the terms with the corresponding operators including necessary shifts of variables to write it in the basis of $a_{k_z}$.

\subsection{CDW contribution to the heat capacity $C_V$}
The formation of the CDW leads to the usual thermdymanic signatures around $T_c$. This is shown explicitly in the inset of Fig. 3 where we have calculated the heat capacity for the model with and without the CDW via
\begin{align}
C_V^\alpha\propto -\frac{\partial U_\alpha}{\partial T}
\    \text{with}  \  \ U_{\alpha}= N_{B} \sum_i E_i^\alpha n_F(E_i^\alpha)
\end{align}
where $E_i^{\alpha}$ with $\alpha=$CDW ($\alpha=0$) are the energy levels with (without) the CDW formation and the multi-label $i$ refers to the LL index, the momentum $k_z$ and the electron- and holelike band index.


\begin{thebibliography}{10}
\bibitem{weyl_elektron_1929}
H. Weyl,
{\em Elektron und Gravitation. I},
\href{http://dx.doi.org/10.1007/bf01339504}{Z. Physik {\bf 56}, 330-352 (1929)}.
\bibitem{volovik_universe_2009}
G. E. Volovik,
{\em The Universe in a Helium Droplet},
\href{http://dx.doi.org/10.1093/acprof:oso/9780199564842.001.0001}{Oxford University Press (2009)}.
\bibitem{murakami_phase_2007}
S. Murakami,
{\em Phase transition between the quantum spin Hall and insulator phases in 3D: emergence of a topological gapless phase},
\href{http://iopscience.iop.org/1367-2630/9/9/356}{New Journal of Physics {\bf 9}, 356 (2007)}.
\bibitem{wan_topological_2011}
X. Wan, A. M. Turner, A. Vishwanath and S. Y. Savrasov,
{\em Topological semimetal and Fermi-arc surface states in the electronic structure of pyrochlore iridates},
\href{http://link.aps.org/doi/10.1103/PhysRevB.83.205101}{Phys. Rev. B {\bf 83}, 205101 (2011)}.
\bibitem{burkov_weyl_2011}
A. A. Burkov and L. Balents,
{\em Weyl Semimetal in a Topological Insulator Multilayer},
\href{http://link.aps.org/doi/10.1103/PhysRevLett.107.127205}{Phys. Rev. Lett. {\bf 107}, 127205 (2011)}.
\bibitem{xu_discovery-taas_2015}
S. Xu, I. Belopolski, N. Alidoust, M. Neupane, G. Bian, C. Zhang, R. Sankar, G. Chang, Z. Yuan, C. Lee, S. Huang, H. Zheng, J. Ma, D. S. Sanchez, B. Wang, A. Bansil, F. Chou, P. P. Shibayev, H. Lin, S. Jia and M. Z. Hasan,
{\em Discovery of a Weyl fermion semimetal and topological Fermi arcs},
\href{http://science.sciencemag.org/content/349/6248/613}{Science {\bf 349}, 613--617 (2015)}.
\bibitem{lv_discovery_2015}
B. Q. Lv, H. M. Weng, B. B. Fu, X. P. Wang, H. Miao, J. Ma, P. Richard, X. C. Huang, L. X. Zhao, G. F. Chen, Z. Fang, X. Dai, T. Qian and H. Ding,
{\em Experimental Discovery of Weyl Semimetal TaAs},
\href{http://link.aps.org/doi/10.1103/PhysRevX.5.031013}{Phys. Rev. X {\bf 5}, 031013 (2015)}.
\bibitem{lu_experimental_2015}
L. Lu, Z. Wang, D. Ye, L. Ran, L. Fu, J. D. Joannopoulos and M. Solja\v{c}i\'{c},
{\em Experimental observation of Weyl points},
\href{http://dx.doi.org/10.1126/science.aaa9273}{Science {\bf 349}, 622-624 (2015)}.
\bibitem{huang_observation_2015}
X. Huang, L. Zhao, Y. Long, P. Wang, D. Chen, Z. Yang, H. Liang, M. Xue, H. Weng, Z. Fang and et al.,
{\em Observation of the Chiral-Anomaly-Induced Negative Magnetoresistance in 3D Weyl Semimetal TaAs},
\href{http://dx.doi.org/10.1103/physrevx.5.031023}{Physical Review X {\bf 5}, 031023 (2015)}.
\bibitem{zhang_signatures_2016}
C. Zhang, S. Xu, I. Belopolski, Z. Yuan, Z. Lin, B. Tong, G. Bian, N. Alidoust, C. Lee, S. Huang and et al.,
{\em Signatures of the Adler-Bell-Jackiw chiral anomaly in a Weyl fermion semimetal},
\href{http://dx.doi.org/10.1038/ncomms10735}{Nat Comms {\bf 7}, 10735 (2016)}.
\bibitem{Moll2016}
J. G. Ashvin
and J. Analytis,
{\em Transport evidence for Fermi-arc-mediated chirality transfer in the Dirac semimetal Cd3As2},
\href{http://dx.doi.org/10.1038/nature18276}{Nature {\bf 535}, 266-270 (2016)}.
\bibitem{soluyanov_type-ii_2015}
A. A. Soluyanov, D. Gresch, Z. Wang, Q. Wu, M. Troyer, X. Dai and B. A. Bernevig,
{\em Type-II Weyl semimetals},
\href{http://dx.doi.org/10.1038/nature15768}{Nature {\bf 527}, 495-498 (2015)}.

\bibitem{bergholtz_topology_2015}
E. J. Bergholtz, Z. Liu, M. Trescher, R. Moessner and M. Udagawa,
{\em Topology and Interactions in a Frustrated Slab: Tuning from Weyl Semimetals to $\mathcal{C}>1$ Fractional Chern Insulators},
\href{http://link.aps.org/doi/10.1103/PhysRevLett.114.016806}{Phys. Rev. Lett. {\bf 114}, 016806 (2015)}.
\bibitem{xu_structured_2015}
Y. Xu, F. Zhang and C. Zhang,
{\em Structured Weyl Points in Spin-Orbit Coupled Fermionic Superfluids},
\href{http://dx.doi.org/10.1103/physrevlett.115.265304}{Phys. Rev. Lett. {\bf 115}, 265304 (2015)}.
\bibitem{goerbig_tilted_2008}
M. O. Goerbig, J. N. Fuchs, G. Montambaux and F. Piechon,
{\em Tilted anisotropic Dirac cones in quinoid-type graphene and alpha-(BEDT-TTF)$_2I_3$},
\href{http://link.aps.org/doi/10.1103/PhysRevB.78.045415}{Phys. Rev. B {\bf 78}, 045415 (2008)}.
\bibitem{trescher_quantum_2015}
M. Trescher, B. Sbierski, P. W. Brouwer and E. J. Bergholtz,
{\em Quantum transport in Dirac materials: Signatures of tilted and anisotropic Dirac and Weyl cones},
\href{http://dx.doi.org/10.1103/physrevb.91.115135}{Phys. Rev. B {\bf 91}, 115135 (2015)}.
\bibitem{ali_large_magnetoresistance_wte2_2014}
M. N. Ali, J. Xiong, S. Flynn, J. Tao, Q. D. Gibson, L. M. Schoop, T. Liang, N. Haldolaarachchige, M. Hirschberger, N. P. Ong and et al.,
{\em Large, non-saturating magnetoresistance in WTe2},
\href{http://dx.doi.org/10.1038/nature13763}{Nature {\bf 514}, 205-208 (2014)}.
\bibitem{Autes2016}
{G. Aut\`es, D. Gresch, M. Troyer, A. A. Soluyanov and O. V. Yazyev,}
{\em Robust Type-II Weyl Semimetal Phase in Transition Metal Diphosphides $X{\mathrm{P}}_{2}$ ($X=\mathrm{Mo}$, W)},
\href{https://link.aps.org/doi/10.1103/PhysRevLett.117.066402}{Phys. Rev. Lett. {\bf 117}, 066402 (2016)}.
\bibitem{Kumar2017}
N. Kumar, Y. Sun, K. Manna, V. Suess, I. Leermakers, O. Young, T. Foerster, M. Schmidt, B. Yan, U. Zeitler, C. Felser and C. Shekhar
{\em Extremely high magnetoresistance and conductivity in the type-II Weyl semimetal WP$_2$},
\href{https://arxiv.org/abs/1703.04527}{arxiv:1703.04527 (2017)}.
\bibitem{belopolski_discovery_2016}
I. Belopolski, D. S. Sanchez, Y. Ishida, P. Xingchen
and Yu, S. Xu, G. Chang, H. Tay-Rong
and Zheng, N. Alidoust, G. Bian, M. Neupane, C. Shin-Ming
and Lee, Y. Song, H. Bu, G. Wang, S. Li, H. Goki
and Jeng, T. Kondo, H. Lin, Z. Liu, F. Song, S. Shin and M. Z. Hasan,
{\em Discovery of a new type of topological Weyl fermion semimetal state in Mo$_x$W$_{1-x}$Te$_2$},
\href{https://doi.org/10.1038/ncomms13643}{Nature Comm. {\bf 7}, 13643 (2016)}.
\bibitem{Udagawa2017}
M. Udagawa and E. J. Bergholtz,
{\em Field-Selective Anomaly and Chiral Mode Reversal in Type-II Weyl Materials},
\href{https://link.aps.org/doi/10.1103/PhysRevLett.117.086401}{Phys. Rev. Lett. {\bf 117}, 086401 (2016)}.
\bibitem{Tchoumakov2017}
S. Tchoumakov, M. Civelli and M. O. Goerbig,
{\em Magnetic-Field-Induced Relativistic Properties in Type-I and Type-II Weyl Semimetals},
\href{https://link.aps.org/doi/10.1103/PhysRevLett.117.086402}{Phys. Rev. Lett. {\bf 117}, 086402 (2016)}.
\bibitem{Yu2017}
Z. M. Yu, Y. Yao and S. A. Yang,
{\em Predicted Unusual Magnetoresponse in Type-II Weyl Semimetals},
\href{https://link.aps.org/doi/10.1103/PhysRevLett.117.077202}{Phys. Rev. Lett. {\bf 117}, 077202 (2016)}.
\bibitem{Meng2016}T. Meng, A.G. Grushin, K. Shtengel, and J. H. Bardarson, {\em Theory of a 3+1D fractional chiral metal: Interacting variant of the Weyl semimetal}, \href{https://journals.aps.org/prb/abstract/10.1103/PhysRevB.94.155136}{Phys. Rev. B {\bf 94}, 155136 (2016)}.
\bibitem{Wei2012}
H. Wei, S. P. Chao and V. Aji,
{\em Excitonic Phases from Weyl Semimetals},
\href{https://link.aps.org/doi/10.1103/PhysRevLett.109.196403}{Phys. Rev. Lett. {\bf 109}, 196403 (2012)}.
\bibitem{Wang2013}
Z. Wang and S. C. Zhang,
{\em Chiral anomaly, charge density waves, and axion strings from Weyl semimetals},
\href{https://link.aps.org/doi/10.1103/PhysRevB.87.161107}{Phys. Rev. B {\bf 87}, 161107 (2013)}.
\bibitem{Laubach2016}
M. Laubach, C. Platt, R. Thomale, T. Neupert and S. Rachel,
{\em Density wave instabilities and surface state evolution in interacting Weyl semimetals},
\href{https://link.aps.org/doi/10.1103/PhysRevB.94.241102}{Phys. Rev. B {\bf 94}, 241102 (2016)}.
\bibitem{Roy2017}
B. Roy, P. Goswami and V. Juri\ifmmode \check{c}\else \v{c}\fi{}i\ifmmode \acute{c}\else \'{c}\fi{},
{\em Interacting Weyl fermions: Phases, phase transitions, and global phase diagram},
\href{https://link.aps.org/doi/10.1103/PhysRevB.95.201102}{Phys. Rev. B {\bf 95}, 201102 (2017)}.
\bibitem{wang_topological_2016}
Y. Wang and P. Ye,
{\em Topological density-wave states in a particle-hole symmetric Weyl metal},
\href{https://link.aps.org/doi/10.1103/PhysRevB.94.075115}{Phys. Rev. B {\bf 94}, 075115 (2016)}.
\bibitem{Yang2011}
K. Y. Yang, Y. M. Lu and Y. Ran,
{\em Quantum Hall effects in a Weyl semimetal: Possible application in pyrochlore iridates},
\href{https://link.aps.org/doi/10.1103/PhysRevB.84.075129}{Phys. Rev. B {\bf 84}, 075129 (2011)}.
\bibitem{Roy2015}
B. Roy and J. D. Sau,
{\em Magnetic catalysis and axionic charge density wave in Weyl semimetals},
\href{https://link.aps.org/doi/10.1103/PhysRevB.92.125141}{Phys. Rev. B {\bf 92}, 125141 (2015)}.
\bibitem{zhang_transport_2017}
X.T. Zhang and R. Shindou,
{\em Transport properties of density wave phases in three-dimensional metals and semimetals under high magnetic field},
\href{https://doi.org/10.1103/physrevb.95.205108}{Phys. Rev. B {\bf 95}, 205108 (2017)}.
\bibitem{sun_helical_2015}
X.Q. Sun, S. Zhang and Z. Wang,
{\em Helical Spin Order from Topological Dirac and Weyl Semimetals},
\href{https://doi.org/10.1103/physrevlett.115.076802}{Phys. Rev. Lett. {\bf 115}, 076802 (2015)}.
\bibitem{Note1}
See Supplemental Material for details of the mean field calculation and additional information on susceptibility and heat capacity.
\bibitem{Chubukov2008}
A. V. Chubukov, D. V. Efremov and I. Eremin,
{\em Magnetism, superconductivity, and pairing symmetry in iron-based superconductors},
\href{https://link.aps.org/doi/10.1103/PhysRevB.78.134512}{Phys. Rev. B {\bf 78}, 134512 (2008)}.
\bibitem{Goerbig_RMP}
M. O. Goerbig, {\em Electronic properties of graphene in a strong magnetic field}, 
\href{https://link.aps.org/doi/10.1103/RevModPhys.83.1193}{Rev. Mod. Phys. {\bf 83}, 1193 (2011)}.
\bibitem{goerbig_competition_2004}
M. O. Goerbig, P. Lederer and C. Morais Smith,
{\em Competition between quantum-liquid and electron-solid phases in intermediate Landau levels},
\href{http://dx.doi.org/10.1103/physrevb.69.115327}{Phys. Rev. B {\bf 69}, 115327 (2004)}.
\bibitem{fukuyama_two-dimensional_1979}
H. Fukuyama, P. M. Platzman and P. W. Anderson,
{\em Two-dimensional electron gas in a strong magnetic field},
\href{http://dx.doi.org/10.1103/physrevb.19.5211}{Phys. Rev. B {\bf 19}, 5211-5217 (1979)}.

\bibitem{Knolle_2015}
J. Knolle, N. R. Cooper, 
{\em Quantum Oscillations without a Fermi Surface and the Anomalous de Haas-van Alphen Effect},
\href{https://journals.aps.org/prl/abstract/10.1103/PhysRevLett.115.146401}{Phys. Rev. Lett. {\bf 115}, 146401 (2015)}.

\bibitem{yoshioka_electronic_1981}
D. Yoshioka and H. Fukuyama,
{\em Electronic Phase Transition of Graphite in a Strong Magnetic Field},
\href{https://doi.org/10.1143/jpsj.50.725}{Journal of the Physical Society of Japan {\bf 50}, 725--726 (1981)}.
\bibitem{fauque_two_2013}
B. Fauqu\'e, D. LeBoeuf, B. Vignolle, M. Nardone, C. Proust and K. Behnia,
{\em Two Phase Transitions Induced by a Magnetic Field in Graphite},
\href{https://link.aps.org/doi/10.1103/PhysRevLett.110.266601}{Phys. Rev. Lett. {\bf 110}, 266601 (2013)}.
\bibitem{leboeuf_thermodynamic_2017}
D. LeBoeuf, C. W. Rischau, G. Seyfarth, R. K\"uchler, M. Berben, S. Wiedmann, W. Tabis, M. Frachet, K. Behnia and B. Fauqu\'e,
{\em Thermodynamic signatures of the field-induced states of graphite},
\href{http://arxiv.org/abs/1705.07056v1}{arxiv:1705.07056v1 (2017)}.
\bibitem{DDSarma_2017}
S. Thirupathaiah, Rajveer Jha, Banabir Pal, J. S. Matias, P. Kumar Das, P. K. Sivakumar, I. Vobornik, N. C. Plumb, M. Shi, R. A. Ribeiro, D. D. Sarma,
{\em What Makes the Extremely Large Magnetoresistance of MoTe$_2$ Unsaturated},
\href{https://arxiv.org/abs/1705.07217}{arxiv:1705.07217 (2017)}.

\bibitem{Bernevig2007}
B. A. Bernevig, T. L. Hughes, S. Raghu and D. P. Arovas,
{\em Theory of the Three-Dimensional Quantum Hall Effect in Graphite},
\href{https://link.aps.org/doi/10.1103/PhysRevLett.99.146804}{Phys. Rev. Lett. {\bf 99}, 146804 (2007)}.
\bibitem{McKernan1995}
S. K. McKernan, S. T. Hannahs, U. M. Scheven, G. M. Danner and P. M. Chaikin,
{\em Competing Instabilities and the High Field Phases of $(\mathrm{TMTSF}{)}_{2}$ ${\mathrm{ClO}}_{4}$},
\href{https://link.aps.org/doi/10.1103/PhysRevLett.75.1630}{Phys. Rev. Lett. {\bf 75}, 1630--1633 (1995)}.
\bibitem{Balicas1995}
L. Balicas, G. Kriza and F. I. B. Williams,
{\em Sign Reversal of the Quantum Hall Number in (TMTSF${)}_{2}$P${\mathrm{F}}_{6}$},
\href{https://link.aps.org/doi/10.1103/PhysRevLett.75.2000}{Phys. Rev. Lett. {\bf 75}, 2000--2003 (1995)}.
\bibitem{bychkov_two-dimensional_1983}
Y. A. Bychkov and E. Rashba,
{\em Two-dimensional electron-hole system in a strong magnetic field: biexcitons and charge-density waves},
{Zh. Eksp. Teor. Fiz. {\bf 85}, 1826-1846 (1983)}.
\end{thebibliography}
\end{document}